\documentclass[12pt]{article}

\usepackage[dvips]{graphics}

\begin{document}
\title{Nonlinearity in the dynamics of photoinduced nucleation process}
\author{Kunio Ishida\\
Corporate Research and Development Center, Toshiba Corporation\\
 1 Komukaitoshiba-cho, Saiwai-ku, Kawasaki 212-8582, Japan\\
and\\
Keiichiro Nasu\\
Solid State Theory Division, Institute of Materials Structure Science, KEK,\\
Graduate University for Advanced Study, and CREST JST,\\ 
1-1 Oho, Tsukuba, Ibaraki 305-0801, Japan}
\date{}

\maketitle

\begin{abstract}
Nonlinear dynamics of photoinduced cooperative phenomena is studied by numerical calculations on a model of molecular crystals.
We found that the photoinduced nucleation process is triggered only when certain amount of excitation energy
is supplied in a narrow part of the system, i.e., there exists a smallest cluster of excited 
molecules which makes the nucleation possible.
As a result, the portion of the cooperatively converted molecules is nonlinearly dependent on the photoexcitation
strength, which has been observed in various materials.
\end{abstract}

\clearpage

The discovery of the photoinduced cooperative phenomena has attracted our
attention to new aspects of the nonequilibrium dynamics of excited states in condensed matter\cite{nasu}.
They have mentioned that the dynamics of photoinduced domain
growth is a key to understand the physics of the photoinduced cooperativity,
and many experimental and theoretical studies have been performed
to reveal the mechanism of these phenomena\cite{ct,pda,spin,binuclear,letard,iwai,ogawa,nasu2,ishida2,ishida3}.
Above all, interdomain interaction and/or nonlinearity in the pattern
formation process of photoinduced domains is important to  understand their dynamical properties.

As for the theoretical study of these phenomena, we have proposed a model of molecular crystals to describe the 
initial nucleation processes of photoinduced domains\cite{ishida2,ishida3}.
We found that an initially excited molecule becomes a nucleus of a
photoinduced domain, and that its growth dynamics is
understood by the numerical solutions of the time-dependent Schr\"odinger equation.
Although we discussed the interdomain interactions in some
special cases\cite{ishida2}, no microscopic theories on the nonlinearity of the domain growth dynamics 
have been obtained so far.

Nonlinear dynamics of spatio-temporal pattern formation
has been extensively studied to understand the various aspects of nonequilibrium phenomena
in wide scale-range, e.g., from nanoscale of the Belousov-Zhabotinsky
reaction or the phase separation dynamics in the kinetic Ising model\cite{ising} to the astronomical
scale of cosmic structure formation\cite{cosmic}.
It has been pointed out that the density fluctuation of
relevant physical properties is a ``seed'' of growing patterns,
and that the initial density distribution determines their complicated structure.
Hence, in analogy to them, it is required to study the domain growth dynamics in the presence of
excitation density fluctuation in order to clarify and understand the
nonlinear nature of the photoinduced cooperativity.

In this Letter, we study the photoinduced nucleation processes focusing on the relation 
between the excited energy distribution and the pattern formation dynamics.
We employ a model of molecules arrayed on a square lattice\cite{ishida2,ishida3},
and show that the spatial fluctuation of excitation density plays 
a key role in the growth/extinction processes of photoinduced
domains as in the case of various types of phase separation dynamics\cite{dcp}.

To describe the present model, we first point out that one of the elementary processes of the photoinduced nucleation in this
system is the nonadiabatic transition between the ground and
the excited electronic states in each molecule.
Electrons relevant to this process are assumed to be
localized in each molecule, and two electronic levels corresponding to
the ground and excited electronic states are taken into account per molecule.
The diabatic potential energy surfaces with a single relevant
vibration mode are assumed in each molecule which cross with each other and
the nonadiabaticity in the dynamics is taken into account via ``spin-flip''
interaction between two electronic states as in the studies of
the nonadiabatic transitions in typical organic molecules\cite{mol}.
As for the intermolecular interaction, we take into account the bilinear coupling terms
between distortion of adjacent molecules and the dipole-dipole interaction between excited state
electrons of which the strength varies with the molecular distortion.
Furthermore, the interaction which describes the molecular distortion induced by
the excited state electrons in the adjacent molecules is also considered.
Hence, the Hamiltonian in the present study is described by:
\begin{eqnarray}
{\cal H} & = & \sum_{\vec{r}} \left \{\frac{p_{\vec{r}}^2}{2}+ \frac{\omega^2 u_{\vec{r}}^2}{2}+ ( \sqrt{2\hbar \omega^3}sq_{\vec{r}}+ \varepsilon \hbar \omega + s^2 \hbar \omega ) \hat{n}_{\vec{r}}+\lambda \sigma_x^{\vec{r}} \right \} \nonumber \\
& + &\sum_{\langle \vec{r},\vec{r'}\rangle} [ \alpha \omega^2
     (u_{\vec{r}}-\beta\hat{n}_{\vec{r}})(u_{\vec{r'}}-\beta\hat{n}_{\vec{r'}}) - \{ V - W (u_{\vec{r}}+u_{\vec{r'}}) \} \hat{n}_{\vec{r}} \hat{n}_{\vec{r'}} ],
\label{ham}
\end{eqnarray}
where $p_{\vec{r}}$ and $u_{\vec{r}}$ are the momentum and coordinate
operators for the vibration mode of a molecule at site $\vec{r}$, respectively.
The electronic states at site $\vec{r}$ are denoted by $|\downarrow \rangle_{\vec{r}}$ (ground
state) and $|\uparrow \rangle_{\vec{r}}$ (excited state) and
$\sigma_i^{\vec{r}}\ (i=x,y,z)$ are the Pauli matrices which act only on
the electronic states of the molecule at site $\vec{r}$.
$\hat{n}_{\vec{r}}$ denotes the density of the electron in $|\uparrow
\rangle_{\vec{r}}$ which is rewritten as $\hat{n}_{\vec{r}}=\sigma_z^{\vec{r}}+1/2$.
The second sum which gives the intermolecular interaction is taken over
all the pairs on nearest neighbor sites, and the vibrational period of an individual molecule is denoted by $T=2\pi/\omega$.
The vibration modes are quantized in order to describe the
nonadiabatic transition between potential energy surfaces rigorously.

We chose the values of the parameters as:
$\varepsilon=2.3$, $s=1.4$, $V=1.1$, $W=0.2$, $\alpha=0.1$, $\beta=0.2$, and $\lambda=0.2$.
Although those values are typical for organic molecules as for
electron-vibration coupling\cite{wp} and  the intermolecular Coulomb interaction,
we mention that the other parameters are not easy to determine their values either from
theoretical calculations or experimental results.

The numerical solutions of the time-dependent Schr\"odinger equation for the
Hamiltonian (\ref{ham}) were obtained by the Runge-Kutta method with
various initial conditions.
In each series of simulations, the dynamics of the system on a $96\times96$ lattice 
with periodic boundary conditions was calculated.
We also applied a mean-field approximation in which the contribution of the wavefunction at the
nearest neighbor sites is substituted by the average value with
respect to the wavefunction $|\Phi(t) \rangle$, which is also quite suitable
for large scale computing with a grid environment, for example.
The details of the model and the calculation method are described in Ref.\ \cite{ishida3}.

\begin{figure}[htb]
\begin{center}
\scalebox{0.5}{\includegraphics*{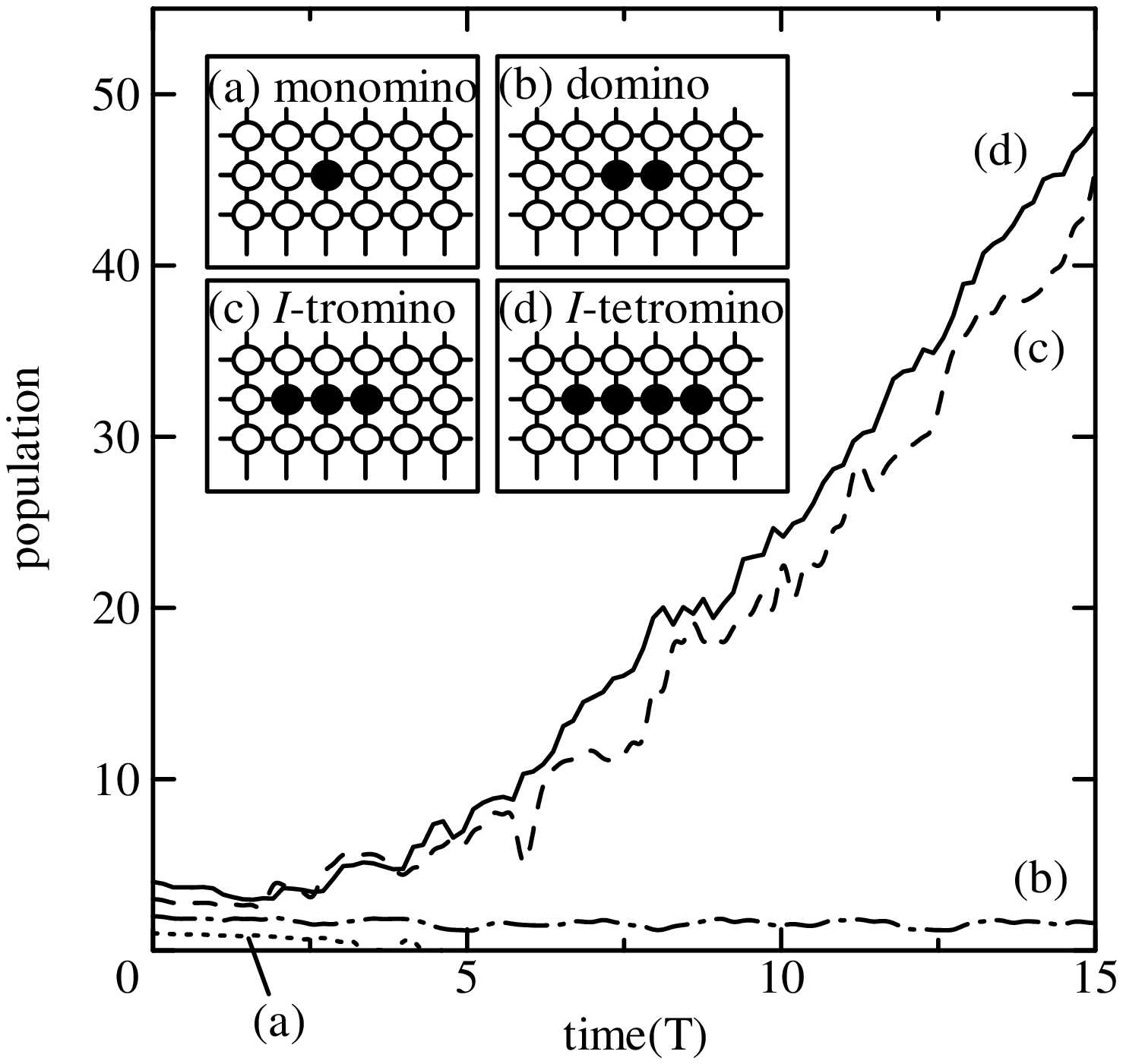}}
\caption{Population of the excited state molecule $N(t)$ as functions
  of time. Inset: initial configurations of excited molecules.
  the filled circles and the open circles correspond to the molecules in
  the Franck-Condon state and the ground state, respectively.}
\label{population}
\end{center}
\end{figure}

We have pointed out that the population of the excited electronic state
$\tilde{N}(\vec{r},t)=\langle \Phi(t)|\hat{n}_{\vec{r}}| \Phi(t) \rangle$ 
is suitable to discuss the dynamics of the spatial patterns of the photoinduced domains.
In particular, the sum of the excited state population $N(t)=\sum_{\vec{r}}\tilde{N}(\vec{r},t)$
gives a measure to estimate the size of the photoinduced domain\cite{ishida3},
which is shown in Fig.\ \ref{population} as functions of time for several initial conditions.
The dotted line in Fig.\ \ref{population} shows that an isolated excited molecule does
not trigger the nucleation processes but loses the excitation energy
after a few periods of the molecular vibration mode.
When two excited molecules are placed on the nearest neighbor sites, the potential energy barrier between $|\downarrow \rangle_{\vec{r}}$
and $|\uparrow \rangle_{\vec{r}}$ in these molecules are still too high to make the
electronic state conversion possible. 
Thus no domain growth occurs though these molecules do not go back to
the ground state within the simulation time(see the dot-dashed line in Fig.\ \ref{population}).
On the contrary, when the initially excited molecules form an $I$-tromino
or an $I$-tetromino, 
$N(t)$ increases as the excitation energy is transferred to the other molecules,
and a photoinduced domain grows as shown in the solid line and the dashed line in Fig.\ \ref{population}.

The above results clearly show that there exists a smallest cluster of excited molecules which leads to the domain growth after photoexcitation, i.e.,
a certain amount of excitation energy is required to be initially concentrated
in a narrow spatial area of the system.
The dynamics of the photoexcited states differs qualitatively from
each other as
the initial state is varied, and, in particular, the size of
the domain is not simply determined by the value of $N(0)$.
To be more precise, the domain growth dynamics reflects the symmetry of the initial
configuration as well as the concentration of excitation energy.
A taxonomic study of the domain growth dynamics with respect to the
configuration of initial excited molecules will be given elsewhere.

Although the size of the smallest cluster for the domain growth
depends on the values of the parameters, we also stress that our simple
model which consists only of localized electrons and molecular vibration
modes is sufficient to discuss the nonlinearity of the nucleation process.

\begin{figure}[htb]
\begin{center}
\scalebox{0.36}{\includegraphics*{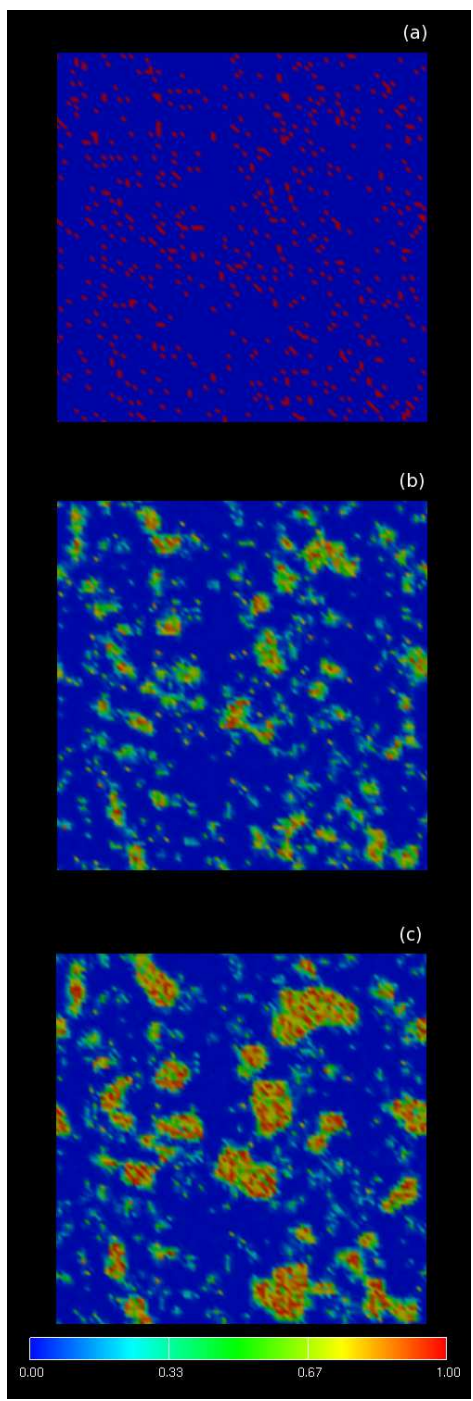}}
\caption{Gradation maps of excited state population $\tilde{N}(\vec{r},t)$on a 96$\times$96 lattice for
  $\rho=0.0625$ for (a)$t=0$, (b) $t=7.5T$, and (c) $t=15T$. }
\label{random}
\end{center}
\end{figure}

We extended our calculations to the cases where initially excited
molecules are distributed at random, and studied the pattern formation dynamics
of excited-state domains during the initial  nucleation processes in
coherent regime.
In this case, the excitation ratio $\rho$, defined by the ratio of the number of the initially excited molecules
 to the total number of the molecules, is a relevant parameter.
Figure \ref{random} shows the gradation maps of $\tilde{N}(\vec{r},t)$ for
$t=0$, $t=7.5T$, and $t=15T$ for $\rho = 0.0625$.
At $t=0$ the distribution of the excited molecules is not 
uniform, and thus the fluctuation of $N(\vec{r},t)$ is present.
When excited molecules are densely concentrated in certain part of
the system, the molecules around them are able to overcome
the potential energy barrier to make the electronic state conversion, 
and thus the excited-state domain starts to grow there.
On the contrary, when the density of the excited molecules is
not sufficiently high for the domain growth, the excitation energy is released
to the other molecules through the vibrational coupling $\alpha$, and
thus the excited molecules return to the ground state.
As a result, we obtain islands of photoinduced domains shown in Fig.\
\ref{random}-(b) around the parts of the system 
which are initially supplied with higher excitation energy.
These islands merge with each other to make larger ones as
shown in Fig.\ \ref{random}-(c), and $N(t)$ continues to increase with time.
Thus, the density fluctuation of the initially excited
molecules reflects the structure of the photoinduced domains,
and {\it vice versa}.
We note that these results directly reflects the discussion on the
smallest cluster for the domain growth shown previously.

\begin{figure}[htb]
\begin{center}
\scalebox{0.5}{\includegraphics*{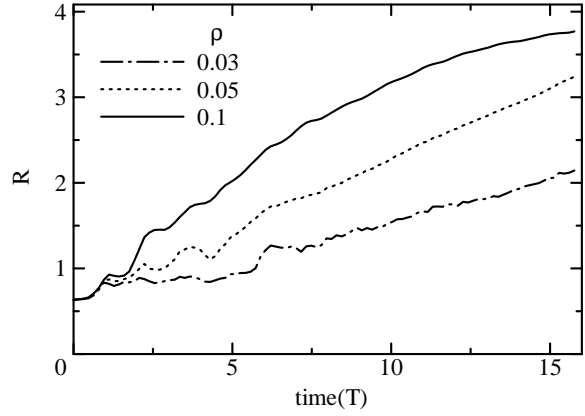}}
\caption{Correlation length $R$ in units of the lattice constant for $\rho=0.03$,
  0.05, and 0.1 as functions of time.}
\label{radius}
\end{center}
\end{figure}

We also calculated the typical correlation length $R$ for the photoinduced domains.
To obtain $R$ we calculated the two-point correlation function 
\begin{equation}
C(|\vec{r}|,t) = \sum_{\vec{r'}}\left (  \langle
\hat{n}_{\vec{r}+\vec{r'}}\hat{n}_{\vec{r'}} \rangle - \langle
\hat{n}_{\vec{r}+\vec{r'}}\rangle \langle \hat{n}_{\vec{r'}} \rangle \right ),
\end{equation}
where $\langle .. \rangle$ denotes the average over 64 series of
simulation for a fixed value of $\rho$.
In each series of simulation, the initial excited molecules are
in different configurations.
Then we define $R$ by the relation $C(R,t) = C(0,t)/e$, and $R$ in units of the lattice constant
is shown in Fig.\ \ref{radius} for $\rho = 0.03$, 0.05, and 0.1.
All of the three lines in Fig.\ \ref{radius} show that $R$ starts to increase
in the same manner, since $R$ for $t \sim 0$ is determined by the
individual clusters of excited molecules.
Although Fig.\ \ref{radius} shows that $R$ is proportional to $t$ for 
small values of $r$ as in the case of a single domain\cite{ishida3}, $R$ increases more slowly for $\rho =0.1$.
This difference is due to the interference between growing domains, i.e., a domain disturbs the growth of the other ones
after they share perimeters with each other.
Hence, the increase rate of $R$ slows down and it seems to behave as
$t^\alpha$ for $\alpha <1$ as in the diffusive domain growth\cite{ising}, for example.
However, we stress that the dynamics considered in the present study is always in coherent regime, and thus the slowdown of the growth rate 
is, as it were,  ``false'' $t$-dependence of $R$.
Such a behavior of $R$ should be discriminated from the diffusive domain growth in the systems which belong to a different universality class.
In particular, since these properties reflects on the structure factor which is the Fourier transformation of $C(|\vec{r}|,t)$,
 they can be distinguished from each other experimentally by varying the excitation ratio.

\begin{figure}[htb]
\begin{center}
\scalebox{0.45}{\includegraphics*{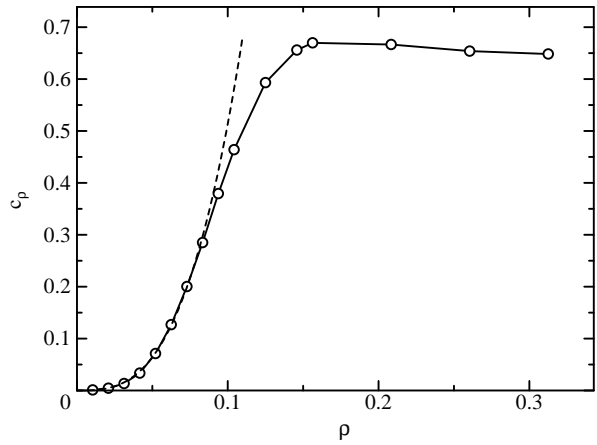}}
\caption{Conversion rate $c_\rho$ as a function of excitation ratio $\rho$. The dashed line which is proportional to $\rho^3$ is drawn as a guide for the eyes.}
\label{conv}
\end{center}
\end{figure}

We also found that the size of the photoinduced domains nonlinearly depends
on the excitation ratio $\rho$.
Figure \ref{conv} shows the conversion rate $c_\rho$ defined by
$c_\rho=N(t=15T)/M$ as a function of $\rho$, where $M$ is the number of 
molecules and is 9216(=96$^2$) in the present calculations.
To obtain these results we calculated the average value of $c_\rho$ for
each value of $\rho$ over 64 series of simulations as in the
calculation of $R$.

Figure \ref{conv} shows that $c_\rho$ depends on $\rho$ as $\sim \rho^3$ in the
dilute limit ($\rho \sim 0$), and deviates from $\rho^3$ for $\rho > 0.1$.
This feature reflects the size of the smallest cluster which enables
domain growth.
With a fixed value of $\rho$, the smallest clusters for domain growth (a tromino) 
appear in the initial state with a probability proportional to $\rho^3$ in the dilute limit.
Hence, only a portion  of the initially excited molecules contributes
to the domain growth.
As we can neglect the interference between domains for $\rho \sim 0$, the
number of converted molecules is proportional to $\rho^3$ in this case.

As $\rho$ increases, the growing domains interfere with each other
and the growth rate becomes lower as shown in Fig.\ \ref{radius}.
Thus $c_\rho$ deviates from $\rho^3$ as the value of $\rho$ increases as
shown in Fig.\ \ref{conv}.
In any case, the conversion rate increases as $\rho^m$ where $m$ is the
size of the smallest cluster which triggers the nucleation processes.
If $m$ is experimentally determined through the measurement of optical
properties, for example, we will have a clue to understand the microscopic
mechanism of the elementary process of the domain growth dynamics.

Figure \ref{conv} also shows that $c_\rho/\rho < 7$ in the present case, although some larger values were reported in experimental studies\cite{ct,pda,iwai,spin}.
Larger values of $c_\rho/\rho$ were found in materials close to their
critical temperature, since instability of thermodynamical state of the system enhances the
conversion rate.
Hence, we mention that the present results correspond to
the cases away from  the critical temperature.
Furthermore, we should note that the present calculations are valid before the decoherence of vibrational states takes place,
and that the value of $c_\rho$ shown in Fig.\ \ref{conv} corresponds to that for $t \sim 3$ psec for $T \sim 200$fsec as in typical organic molecules.
We, however, expect that the domains continue to grow after the decoherence occurs,
and thus the experimentally obtained conversion rate cannot be directly
compared with the present results quantitatively.
We stress that the nonlinearity of conversion rate as a function of $\rho$ is essentially understood by the initial process of the domain growth,
and that the present calculations are of importance in order to understand the
microscopic mechanism of the photoinduced cooperativity.

Summarizing the present Letter, we conclude that the nonlinear dynamics of
the photoinduced domain growth is understood by a model of localized
electrons coupled with molecular vibration mode shown by Eq.\ (\ref{ham}).
We found that there exists a smallest cluster of excited
molecules which is required to induce the domain growth, and that the initial nucleation process does not start unless sufficient
energy is concentrated in a narrow area of the system.
When the fluctuation of excitation density realizes the formation of
such clusters, photoinduced nucleation is triggered to form domains afterwards,
and thus the excitation energy fluctuation determines the domain structure.
We also showed that the nonlinearity of the conversion rate as a
function of excitation ratio is understood as a result of the interdomain
interactions besides the above effect.

K. I. is grateful to K. Takaoka, H. Asai, and S. Nunoue for helpful advice. 
This work was supported by the Next Generation Super Computing Project,  
Nanoscience Program, MEXT, Japan, and the numerical calculations were carried out on the computers at the
Research Center for Computational Science, National Institutes of Natural
Sciences.

\end{document}